\documentclass[11pt,twocolumn]{article}

\usepackage{times,amssymb}
\usepackage[bookmarks=true,
bookmarksopen=true,
bookmarksnumbered=true,
colorlinks,
linkcolor=blue]{hyperref}

\title{Geodesic Paths On 3D Surfaces: Survey and Open Problems}
\author{Anil Maheshwari \and Stefanie Wuhrer}
\date{}
\begin{document}
\maketitle

This survey gives a brief overview of theoretically and practically relevant algorithms to compute geodesic paths and distances on three-dimensional surfaces. The survey focuses on polyhedral three-dimensional surfaces.

\section{Introduction}
\label{Introduction}

Finding shortest paths and shortest distances between points on a surface $S$ in three-dimensional space is a well-studied problem in differential geometry and computational geometry. The shortest path between two points on $S$ is denoted a \textit{geodesic path} on the surface and the shortest distance between two points on $S$ is denoted a \textit{geodesic distance}. In this survey, we consider the case where a discrete surface representation of $S$ is given. Namely, $S$ is represented as a polyhedron $P$ in three-dimensional space. Since discrete surfaces cannot be differentiated, methods from differential geometry to compute geodesic paths and distances cannot be applied in this case. However, algorithms from differential geometry can be discretized and extended. Furthermore, the discrete surface can be viewed as a graph in three-dimensional space. Therefore, methods from graph theory and computational geometry have been applied to find geodesic paths and distances on polyhedral surfaces. 

The general problem of computing a shortest path between polyhedral obstacles in 3D is shown to be NP hard by Canny and Reif using reduction from 3-SAT \cite{canny_reif_87_npHard}. Computing a geodesic path on a polyhedral surface is an easier problem and it is solvable in polynomial time.

Computing geodesic paths and distances on polyhedral surfaces is applied in various areas such as robotics, geographic information systems (GIS), circuit design, and computer graphics. For example, geodesic path problems can be applied to finding the most efficient path a robotic arm can trace without hitting obstacles, analyzing water flow, studying traffic control, texture mapping and morphing, and face recognition. A survey related to geodesic paths in two- and higher-dimensional spaces can be found in the Handbook of Computational Geometry \cite{mitchell_survey_shortest_path}.

Note that the geodesic distance between any two points on $P$ can be easily determined if the geodesic path is known by measuring the (weighted) length of the geodesic path. Hence, we will only consider the problem of computing geodesic paths on $P$.

Problems on finding geodesic paths and distances depending on the number of source and destination points have been studied. The three most commonly studied problems are (a) finding the geodesic path from one source vertex $s \in P$ to one destination vertex $d \in P$, (b) finding the geodesic paths from one source vertex $s \in P$ to all destination vertices in $P$, or equivalently, finding the geodesic paths from all source vertices in $P$ to one destination vertex $d \in P$, known as \textit{single source shortest path (SSSP) problem}, and (c) finding the geodesic paths between all pairs of vertices in $P$, known as \textit{all-pairs shortest path (APSP) problem}. 

The algorithms reviewed in this survey are compared by means of the following five categories:
\begin{itemize}
\item Accuracy of the computed geodesic path.
\item Cost metric used to compute the geodesic path. The cost metric can be the Euclidean distance or a weight function (for example when going uphill is more costly than going downhill).
\item Space complexity of the algorithm.
\item Time complexity of the algorithm.
\item Applicability of the algorithm by surveying if the algorithm has been implemented and tested in practice.
\end{itemize}

Approximation algorithms are compared according to their \textit{approximation ratio} (or \textit{approximation factor}) $k$. An algorithm that finds approximations to a geodesic path with approximation ratio $k$ returns a path of length at most $k$ times the exact geodesic path.

To solve the problem of computing geodesic paths on discrete surfaces, two different general approaches can be used. First, the polyhedral surface can be viewed as a graph and algorithms to compute shortest paths on graphs can be extended to find geodesic paths on polyhedral surfaces. Algorithms following this approach are reviewed in Section~\ref{polyhedral}. Second, the polyhedral surface can be viewed as a discretized differentiable surface and algorithms from differential geometry can be extended to find geodesic paths on polyhedral surfaces. Algorithms following this approach are reviewed in Section~\ref{smooth}. At the end of each section, open problems related to the section are summarized.

\section{Graph-Based Algorithms}
\label{polyhedral}

This section reviews algorithms to compute geodesic shortest paths that can be viewed as extensions of graph theoretic algorithms. To obtain a good understanding of the reviewed algorithms, we first review some well-known graph theoretic algorithms. 

Dijkstra proposed an algorithm to solve the SSSP problem on a directed weighted graph $G(V,E)$ with $n$ vertices, $m$ edges, and positive weights \cite{Dijkstra1959}. Dijkstra's algorithm proceeds by building a list of processed vertices for which the shortest path to the source point $s$ is known. The algorithm iteratively decreases estimates on the shortest paths of non-processed vertices, which are stored in a priority queue. In each iteration of the algorithm, the closest unprocessed vertex from the source is extracted from the priority queue and processed by relaxing all its incident edges. The notion of relaxation underlines the analogy between the length of the shortest path and the length of an extended tension spring. When the algorithm starts, the length of the shortest path is overestimated and can be compared to an extended spring. In each iteration, a shorter path is found, which can be compared to relaxing the spring. Although the original implementation requires $O(n^2)$ time, the running time can be decreased to $O(n\log n+m)$ by using Fibonacci heaps. Thorup~\cite{thorup_99} presented an $O(m)$-time algorithm in case where each edge is assigned a positive integer weight. The main idea is to use a hierarchical bucketing structure to avoid the bottleneck caused by sorting the vertices in increasing order from $s$.

The length of a path on $S$ depends on the employed cost metric. Hence, the shortest or geodesic path on $S$ depends on this cost metric. In Section~\ref{euclidean}, geodesic path algorithms with Euclidean cost metric are reviewed. Using the Euclidean cost metric implies that the Euclidean length of the path is used to measure the length of the path. In Section~\ref{weighted}, geodesic path algorithms on weighted surfaces are reviewed. Using a weighted cost metric implies that different faces of $S$ can be weighted differently. Clearly, any algorithm that can solve a shortest path problem using a weighted cost metric can also solve the same problem using the Euclidean cost metric.

\subsection{Euclidean Cost Metric}
\label{euclidean}

When using the Euclidean distance along a polyhedron $P$ as cost metric, shortest paths consist of straight line segments that cross faces of the polyhedron. An approach to compute shortest paths on $P$ aims to compute a superset of all the possible edges of shortest paths on $P$ and to use this information to compute shortest paths. Since all of the algorithms pursuing this strategy are mainly of theoretical interest to establish bounds on the number of possible edges of shortest paths on $P$, they are not discussed in detail in this survey. The algorithm's pursuing this approach are less efficient than the ones surveyed. A good overview of the algorithms finding edge sequences is given by Lanthier \cite[p. 30--35]{lanthier_99_phdThesis}.

We first review algorithms that only operate on the surface of a convex polyhedron. Second, we review algorithms that operate on the surface of any (convex or non-convex) polyhedron.

\subsubsection*{Convex Polyhedra}

This section discusses algorithms that operate on the surface of a convex polyhedron $P$ with $n$ vertices. Shortest paths according to the Euclidean cost metric are considered. 

Sharir and Schorr \cite{sharir_schorr_86_geodesics_convex_polyhedron} proposed an algorithm that computes the exact shortest path between points on the surface of $P$. The proposed algorithm is mainly based on three observations. First, any shortest path intersecting an edge $e$ of $P$ enters and leaves $e$ under the same angle. Second, no shortest path on a convex polygon $P$ can pass through a vertex $p$ of $P$ unless $p$ is the destination or source of the shortest path. Third, if the sequence of edges of $P$ intersected by the shortest path between $s$ and $p$ is known, the shortest path can be computed as the straight line joining $s$ and $p$ after unfolding the faces adjacent to the edge sequence to a plane. The authors aim to subdivide $P$ with respect to a given source point $s$, such that the shortest path from $s$ to any other point in $P$ can be found efficiently. They define \textit{ridge points} $x$ of $P$ as points that have the property that there exists more than one shortest path from $s$ to $x$ and prove that the ridge points can be represented by $O(n^2)$ straight line segments. The algorithm partitions the boundary of $P$ into at most $n$ connected regions called \textit{peels} not containing any vertices or ridge points of $P$. The boundaries of peels contain only ridge points, vertices, and edges of $P$. The algorithm to construct the peels is similar to Dijkstra's graph search algorithm. The peels are then iteratively unfolded to the plane. The algorithm first preprocesses $P$ by constructing the peels with respect to $s$ in $O(n^3 \log n)$ time. The algorithm stores the computed peels in a tree called slice tree that can then be used to determine the shortest path between an arbitrary point on $P$ and $s$ in $O(n)$ time. The slice tree data structure uses $O(n^2)$ space. 

Mount \cite{mount_85_geodesics_convex} improves the algorithm by Sharir and Schorr both in terms of time and space complexity. The main observation by Mount is that the peels defined by Sharir and Schorr can be viewed as Voronoi regions of a point set $R$ containing the planar unfoldings of the source point $s$. Note that $R$ contains at most $n$ points per face of $P$ because there are at most $n$ peels intersecting a face of $P$. Mount observes that the shortest path from $s$ to any point $x$ on $P$ is at most the shortest path from $x$ to any point $r \in R$ plus the distance between $r$ and $s$ along the planar unfolding of the path. This observation depends on the convexity of $P$ and on the fact that all shortest paths unfold to polygonal chains consisting of straight line segments. Mount uses this observation to prove that the Voronoi regions of $R$ are identical to the peels of $P$ with $s$ as source point. An algorithm following the outline of Dijkstra's algorithm is used to compute the point set $R$ and simultaneously, the Voronoi regions of $R$. Using this approach, $P$ can be preprocessed with respect to $s$ in $O(n^2 \log n)$ time and $O(n^2)$ space. The space requirement to store the data structure after building it can be reduced to $O(n \log n)$ by storing $O(n)$ different but similar lists of size $O(n)$ each in an efficient way to avoid redundancy. Note that building the data structure still requires $O(n^2)$ space. The query time to compute a shortest path from an arbitrary point $p \in P$ to $s$ is reduced to $O(k + \log n)$, where $k$ is the number of faces of $P$ intersected by the shortest path by using an output sensitive point location data structure. Mount \cite{mount_86_storing_subdicision_geodesics} reduced the space requirement to build the data structure storing the Voronoi diagram to $O(n \log n)$ by building a hierarchical structure on the intersections between edges of $P$ and geodesic paths starting from $s$. The data structure stores for each edge $e$ of $P$ a tree whose leaves contain the intersections between $e$ and geodesic paths crossing $e$ in order. Common sub-trees of different edges are shared to reduce the space complexity.

To avoid the high time complexity of finding geodesic paths, Hershberger and Suri \cite{hershberger_suri_95_geodesic_convex_approx} propose an algorithm that finds an approximate shortest path between two points on the surface of $P$. The algorithm takes only $O(n)$ time and has an approximation factor of $2$. The main idea of the algorithm is to extend the notion of bounding boxes to a general simplified representation of $P$ and to compute the shortest path between two points on this simplified shape. To compute a shortest path between $s$ and $t$, the faces containing the source and destination points are extended into planes, and at most $O(n)$ different planes are added to the two planes to obtain a wedge. The shortest paths between the two points are computed on each of these wedges in $O(1)$ time as a wedge has constant description size. The shortest path that was found is used to approximate the shortest path on $P$. The algorithm can be extended to approximately solve the SSSP problem in $O(n \log n)$ time. That is, starting from one source point, the algorithm computes approximations with approximation ratio $2$ to all other points on $P$.

Har-Peled et al. \cite{har-peled_sharir_varadarajan_96_approximate_geodesics_convex} extend the algorithm by Hershberger and Suri to obtain an approximation ratio of $(1+\epsilon)$ for $0<\epsilon<1$. The algorithm is based on the approximation scheme by Dudley~\cite{Dudley1974} that approximates the minimum number of sets required to approximate every set as $\epsilon$-approximation. The algorithm by Har-Peled et al. proceeds by expanding $P$ by a factor related to $\epsilon$ and to the approximation obtained by Hershberger and Suri's algorithm. Denote the expanded polygon by $P'$. The shortest path between two vertices on $P$ is approximated on a grid lattice between the boundaries of $P$ and $P'$. Since $P$ is convex and since the path is not in the interior of $P$, the length of the path cannot be shorter than the true shortest path. The path obtained by this method can be projected to $P$ while ensuring that the length of the path does not grow. The running time of the algorithm is $O(n \min(\frac{1}{\epsilon^{1.5}}, \log n)+ \frac{1}{\epsilon^{4.5}} \log \frac{1}{\epsilon})$ and hence depends both on $n$ and $\epsilon$. As the algorithm by Hershberger and Suri, this algorithm can be extended to approximately solve the SSSP problem. The running time of the extended algorithm is $O(\frac{n}{\epsilon^{4.5}} (\log n+\log \frac{1}{\epsilon}))$. Although the theory used by Har-Peled et al. is rather technical, the algorithm itself is simple. Agarwal et al. \cite{agarwal_har-peled_sharir_varadarajan_97_approximating_geodesics} improved the running time of the algorithm to approximate one shortest path by an approximation ratio of $(1+\epsilon)$ by Har-Peled et al. to $O(n \log \frac{1}{\epsilon} + \frac{1}{\epsilon^3})$. This improves the running time of the algorithm to approximately solve the SSSP problem to $O(\frac{n}{\epsilon^3}+\frac{n}{\epsilon^{1.5}} \log n)$. 
Har-Paled~\cite{har-peled_97_convex} presents a further improvement of the running time of this algorithm. After preprocessing the convex polytope in $O(n)$ time, an $(1+\epsilon)$ approximation of the shortest path between two vertices is reported in $O(\frac{\log n}{\epsilon^{1.5}} + \frac{1}{\epsilon^{3}})$ time. This improves the running time of the algorithm to approximately solve the SSSP problem to $O(n(1+\frac{\log n}{\epsilon^{1.5}} + \frac{1}{\epsilon^{3}}))$. 

Recently, Schreiber and Sharir~\cite{schreiber_sharir_2006_optimal_geodesic_convex} proposed an exact solution to the SSSP problem on convex polyhedra in $3$-dimensional space. The algorithm extends Dijkstra's algorithm to allow continuous updates. That is, a wavefront is propagated from the source $s$ along the boundary of $P$ and the wavefront is updated at events that change the topology of the wavefront. Note that a similar technique of continuous Dijkstra updates was used in \cite{mount_85_geodesics_convex}. The general idea of the continuous Dijkstra technique was formally described by Mitchell et al. \cite{mitchell_mount_papadimitriou_87_geodesic} and is reviewed later in this survey. An implicit representation of the solution is computed in optimal time $O(n \log n)$. The implicit representation is stored using $O(n \log n)$ space. Afterwards, the shortest path from the source to any point on $P$ can be reported in $O(\log n+k)$ time, where $k$ is the number of faces of $P$ crossed by the path.

Schreiber~\cite{schreiber_07} extends the previous approach by Schreiber and Sharir~\cite{schreiber_sharir_2006_optimal_geodesic_convex} to so-called \textit{realistic polyhedra}. Realistic polyhedra are defined as three classes of non-convex polyhedra. The first class of realistic polyhedra have a boundary that forms a terrain whose maximal facet slope is bounded by a constant. The second class of realistic polyhedra has the property that each axis parallel square with edge length $l$ that has distance at least $l$ from $P$ is intersected by at most a constant number of faces of $P$. The third class of realistic polyhedra has the property that for each edge $e$ of $P$ of length $\left|e\right|$, there are at most a constant number of faces within shortest path distance $O(\left|e\right|)$.

Agarwal et al.~\cite{agarwal_har-peled_karia_02_implement} propose an algorithm to compute a $(1+\epsilon)$ approximation of the shortest path between two vertices that uses $O(\frac{n}{\sqrt{\epsilon}})$ time and $O(\frac{1}{\epsilon^4})$ space. The approach proceeds by constructing a graph, computing the shortest path on this graph, and projecting the computed graph onto the surface of $P$. Agarwal et al. implemented and tested this algorithm and the algorithm by Hershberger and Suri \cite{hershberger_suri_95_geodesic_convex_approx} for artificial data sets with up to almost $100000$ faces.

The following problems related to computing Euclidean shortest paths on the surface of a convex polyhedron remain unsolved:
\begin{itemize}
\item Can SSSP problem on convex polyhedra be solved in $O(n \log n)$ time using $O(n)$ space~\cite{schreiber_sharir_2006_optimal_geodesic_convex}? 
\item Can an efficient trade off between the query time and the space complexity be established~\cite{schreiber_sharir_2006_optimal_geodesic_convex}?
\end{itemize}

\subsubsection*{General Polyhedra}

This section discusses algorithms that operate on the surface of a polyhedron $P$ with combinatorial complexity $n$. Note that $P$ need not be convex. Shortest paths according to the Euclidean cost metric are considered. The main problem that occurs when allowing non-convex polyhedra is that geodesic paths from $s$ to $t$ on $P$ may pass through a vertex $p$ of $P$. 

O'Rourke et al.~\cite{orourke_suri_booth_85_geodesics} extend the algorithms by Sharir and Schorr~\cite{sharir_schorr_86_geodesics_convex_polyhedron} and Mount~\cite{mount_85_geodesics_convex} to obtain the first algorithm that finds the exact geodesic path between two vertices of an arbitrary polyhedron in polynomial time. Both the source and the destination point of $P$ are considered to be vertices of $P$. The algorithm considers the problem in two steps. First, the straight-line distances between all pairs of vertices of $P$ are found. This is achieved by extending the technique to compute peels in~\cite{sharir_schorr_86_geodesics_convex_polyhedron}. Second, the shortest distance between the source and the destination vertex is found on the graph induced by the vertices of $P$. The algorithm takes $O(n^5)$ time to compute one shortest path on $P$. Since the complexity of the running time is high, the algorithm is irrelevant for practical purposes and has not been implemented.

Mitchell et al.~\cite{mitchell_mount_papadimitriou_87_geodesic} formalize the technique called \textit{continuous Dijkstra} previously used in~\cite{mount_85_geodesics_convex} to find shortest paths from a source point $s$ on the surface of a convex polyhedron. The algorithm traverses the graph induced by $P$ similarly to the graph exploration of a graph $G$ in Dijkstra's algorithm. Edges of $P$ behave like nodes in $G$. Since the distance from $s$ on $P$ to an edge $e$ is not unique, $e$ is labeled by a function describing the distance from $s$ to $e$. The algorithm keeps track of a subdivision of $e$ with the property that for two points $p$ and $q$ in the same region of $e$, the shortest paths from $s$ to $p$ and from $s$ to $q$ pass through the same sequence of vertices and edges of $P$. Mitchell et al. observe that these subdivisions of $e$ resemble the peels used in~\cite{sharir_schorr_86_geodesics_convex_polyhedron}. However, special care needs to be taken when computing this subdivision, since geodesic paths emanating from $s$ can pass through a vertex $p$ of $P$. In this case, $p$ is treated as a \textit{pseudo-source}. The pseudo-source $p$ is labeled by the geodesic distance from $s$ to $p$. For any point $x$ of $P$, the geodesic distance is the minimum of the shortest distance from $s$ to $x$ not passing through a vertex of $P$ and the geodesic distance from the nearest pseudo-source of $P$ to $x$ plus the label of the pseudo-source. This observation allows to compute the subdivision of $s$ and to store for each region of the subdivision the distance to the nearest pseudo-source. For a given source vertex $s$, the algorithm computes a subdivision of $P$ in $O(n^2 \log n)$ time and $O(n^2)$ space. Once the subdivision has been computed, the distance from $s$ to any other point on $P$ can be computed in $O(\log n)$ time. Reporting the shortest path between $s$ and any other point on $P$ takes $O(k+\log n)$ time, where $k$ is the number of faces of $P$ crossed by the shortest path. If the algorithm is initialized with more than one source point, the subdivision obtained after the continuous Dijkstra algorithm ended represents the Voronoi diagram of the source points. Mitchell et al.'s algorithm is of theoretical interest, since the continuous Dijkstra technique can be used with different update schemes to obtain new algorithms, as we saw for convex polyhedra~\cite{schreiber_sharir_2006_optimal_geodesic_convex}. Although the contribution by Mitchell et al. is rather technical, the algorithm is of practical interest as well. Recently, Surazhkhy et al.~\cite{surazhsky_surazhsky_kirsanov_gortler_hoppe_05_geodesics_meshes} implemented and tested the algorithm on data sets obtained using a laser-range scanner. Although the worst-case running time of the algorithm is $O(n^2 \log n)$, Surazhsky et al. found the algorithm's average running time in their experiments to be much lower and suitable for objects with hundreds of thousands of triangles. The exact algorithm by Mitchell et al. is then modified to obtain an algorithm that solves the SSSP problem with approximation ratio $(1+\epsilon)$. Surazhsky et al. derive from their experiments that an average running time of $O(n \log n)$ can be expected in practice for bounded approximations from one source point to all the other points of the mesh.

Chen and Han~\cite{chen_han_92_shortest_path} developed an algorithm to compute geodesic distances from a source point $s$ on a non-convex polyhedron that does not use the continuous Dijkstra technique. The algorithm constructs a tree called \textit{sequence tree} that can be viewed as an extension of the dual graph of the tree containing ridge points used by Sharir and Schorr~\cite{sharir_schorr_86_geodesics_convex_polyhedron} to non-convex polyhedra. In the case of a convex polyhedron, the sequence tree $T$ contains nodes consisting of an edge $e$ of $P$, the image of $s$ in the local coordinate system of the face incident to $e$, and the projection of the image onto $e$. Chen and Han prove that $T$ has a linear number of nodes, contains all of the shortest paths, and can be built in $O(n^2)$ time. For non-convex polyhedra, $T$ contains additional leaves representing pseudo-sources of $P$ (as defined by Mitchell et al.~\cite{mitchell_mount_papadimitriou_87_geodesic}) and the distances of pseudo-sources from $s$. This augmentation of $T$ adds at most $O(n)$ nodes. Hence, the algorithm builds a sequence tree in $O(n^2)$ time and $O(n)$ space. After $T$ was computed, the geodesic distance between $s$ and any point in $P$ can be reported in $O(\log n)$ time. The geodesic path can be reported in $O(\log n+k)$ time, where $k$ is the number of faces of $P$ crossed by the path.
Kavena and O'Rourke~\cite{kaneva_orourke_00_implement_chan_han} implemented and tested the algorithm on synthetic data. The implementation confirms the quadratic time complexity and the linear space complexity in practice. The experiments show that roundoff errors are not a serious problem for this algorithm. Kavena and O'Rourke found the space complexity to be the bottleneck of the algorithm. Data sets with tens of thousands of points were used to test the algorithm.

Har-Peled~\cite{har-peled_98_general} extended his previous approach to compute $(1+\epsilon)$ approximations of geodesic paths on realistic polyhedra~\cite{har-peled_97_convex} to work in case of general polyhedra. Given a source point $s$, the algorithm computes a subdivision of $P$ of size $O\left(\frac{n}{\epsilon}\log\left(\frac{1}{\epsilon}\right)\right)$ in time $O\left(n^2\log n+\frac{n}{\epsilon} \log \left(\frac{1}{\epsilon} \right) \log \left(\frac{n}{\epsilon} \right)\right)$. In the special case of convex polyhedra, the preprocessing time becomes $O\left( \left(\frac{n}{\epsilon}\right)^3 \log\left(\frac{1}{\epsilon}\right) + \frac{n}{\epsilon^{1.5}} \log\left(\frac{1}{\epsilon}\right) \log n \right)$. After this preprocessing step, a $(1+\epsilon)$ approximation of the shortest path between $s$ and any point $p$ on $P$ can be reported in $O\left(\log\left(\frac{n}{\epsilon}\right)\right)$ time. This implies that a $(1+\epsilon)$ approximation to the SSSP problem can be obtained in $O\left(n^2\log n+\frac{n}{\epsilon} \log \left(\frac{1}{\epsilon} \right) \log \left(\frac{n}{\epsilon} \right) + n \log\left(\frac{n}{\epsilon}\right) \right)$ time.

In 1997, Agarwal and Varadarajan~\cite{agarwal_varadarajan_97_approximate_geodesics} proposed an algorithm that answers the question of whether it is possible to compute an approximate shortest path between two points on polyhedra in sub-quadratic time. The proposed algorithm only works for polyhedra of genus zero. Two algorithms using the same general technique were proposed. The first algorithm computes an approximation to the shortest path with approximation ratio $7(1+\epsilon), \epsilon>0$ in $O(n^{\frac{5}{3}} \log^{\frac{5}{3}}n)$ time. The second algorithm takes only $O(n^{\frac{8}{5}} \log^{\frac{8}{5}}n)$ time, but the approximation ratio increases to $15(1+\epsilon), \epsilon>0$. Note that the running times of both algorithms is independent of the choice of $\epsilon$. The main idea of the algorithm is to partition the boundary of the simple polyhedron $P$ into patches of faces of $P$. A graph $G_i$ is constructed on the boundary of each patch $P_i$. The graphs $G_i$ are merged into one graph $G$ and the geodesic paths on $P$ are approximated by the solution of Dijkstra's algorithm on $G$. Although this is the first paper to break the quadratic time complexity, the contribution is mainly of theoretic interest because the algorithm is involved. Hence, the algorithm has not been implemented.

Kapoor~\cite{kapoor_99_geodesics} presents an algorithm that solves the problem of computing the exact shortest path between a pair of points on $P$ in sub-quadratic time. The algorithm follows the continuous Dijkstra technique by Mitchell et al.~\cite{mitchell_mount_papadimitriou_87_geodesic} and propagates a wavefront over the surface of $P$ starting from a source point $s$. The algorithm maintains the wavefront as a collection of circular arcs with centers at $s$ and pseudo-sources of $P$. Furthermore, the algorithm maintains all of the edges of $P$ that have not yet been reached by the wavefront. The algorithm takes $O(n \log^2 n)$ time and $O(n)$ space. According to O'Rourke~\cite{orourke_column}, the details of the algorithm are \textit{``formidable"}. It is therefore not surprising that the algorithm contains some flaws~\cite{schreiber_sharir_2006_optimal_geodesic_convex}.

Kanai and Suzuki~\cite{kanai_suzuki_01_iterative_geodesic_refinement} propose an iterative approximation algorithm to compute the shortest path between pairs of points on $P$. The algorithm is based on Dijkstra's algorithm and iteratively refines the mesh in regions where the path can pass. The refinement proceeds by placing Steiner points on edges of $P$ and to repeat Dijkstra's algorithm on the augmented graph. The user gives two thresholds related to the accuracy of the approximation. The first threshold defines the number of times the algorithm iterates. The second threshold is related to the number of Steiner points placed on an edge of $P$. The authors implement the algorithm and compare it to an implementation of Chen and Han's algorithm. They find their algorithm to outperform Chen and Han's algorithm both in terms of time and space complexity.

The following problems related to computing Euclidean shortest paths on the surface of a possibly non-convex polyhedron remain unsolved:
\begin{itemize}
\item Can the exact shortest path between a pair of vertices on $P$ be computed in $O(n \log n)$ time using $O(n)$ space?
\item Can the SSSP problem be solved in $O(n \log n)$ time and $O(n)$ space?
\end{itemize}

\subsection{Weighted Cost Metric}
\label{weighted}
This section discusses algorithms that operate on the surface of a possibly non-convex polyhedron $P$ with combinatorial complexity $n$ in $3$-dimensional space. Unlike in Section~\ref{euclidean}, the length of the shortest path is not simply measured by its Euclidean length. Instead, a weight $w_i$ is associated with each face $f_i$ of $P$. The length of a path crossing $f_i$ is its Euclidean length multiplied by $w_i$. The weights can be used to model the difficulty of the path. For example, it is harder to walk on an uneven terrain than on an asphalt road. A good overview of algorithms related to weighted shortest paths can be found in Lanthier~\cite{lanthier_99_phdThesis}.

Mitchell and Papadimitriou~\cite{mitchell_papadimitriou_87_weighted_geodesics} present an algorithm to compute the shortest path distance between two arbitrary points in a planar subdivision with $n$ edges. They note that shortest paths obey Snell's Law of refraction at edges of the subdivision. The algorithm is based on this observation and the continuous Dijkstra technique formalized in~\cite{mitchell_mount_papadimitriou_87_geodesic}. Therefore, the authors note that the algorithm can be extended to compute weighted shortest paths on the surface of possibly non-convex polyhedra. The algorithm finds an approximation of the shortest path with approximation ratio $(1+\epsilon)$ using $O(n^8 \log(nN\frac{W}{w\epsilon}))$ time and $O(n^4)$ space, where $N$ is the largest integer coordinate of any vertex in the subdivision and $\frac{W}{w}$ is the ratio between the maximum and the minimum weight. To our knowledge, this algorithm has not been implemented. This is not surprising, since the high time complexity makes the algorithm unsuitable for practical purposes.

Lanthier et al.~\cite{lanthier_maheshwari_sack_97_weighted_geodesics, lanthier_maheshwari_sack_01_weighted_geodesics} developed an approach to construct a graph that can be searched to obtain an approximate shortest path on $P$ by adding Steiner points on each edge of $P$. Without loss of generality, the authors assume $P$ to be triangulated. A total of $O(n^2)$ Steiner points are added, yielding space complexity $O(n^2)$ for all of the algorithms. Four algorithms are presented. The first algorithm computes the shortest path between two arbitrary points on $P$ by finding shortest paths in the graph containing vertices of $P$ and the added Steiner points. The computed shortest path is at most $WL$ longer than the true weighted shortest path on $P$, where $W$ is the maximum weight and where $L$ is the longest edge of $P$. The running time of this algorithm is $O(n^5)$. Second, a faster and less accurate algorithm is presented to compute the shortest path between two arbitrary points on $P$ by computing a spanner on the graph containing the Steiner points and by finding a shortest path on the spanner. The computed shortest path has length at most $\beta(\pi+WL)$, where $\pi$ is the true weighted shortest path on $P$ and $\beta>1$ is a constant. Third, algorithms were presented to process $P$ for queries asking for shortest paths between a fixed source point $s$ in $P$ and an arbitrary point $q$ in $P$. The query time is proportional to $\log n$ and the accuracy of the shortest path. Fourth, algorithms were presented to process $P$ for queries asking for shortest paths between two arbitrary points in $P$. The query time is proportional to $\log n$ and the accuracy of the shortest path. For the query algorithms, time-space trade off schemes are presented. The authors implemented and tested all of the algorithms on both real-life and synthetic data sets. Experiments showed that in practice, much less than $n^2$ Steiner points suffice to yield acceptable results.

Lanthier et al. furthermore present an algorithm that runs in $O(n\log n)$ time that computes a shortest path withing a factor of $\left(1+\frac{2}{\sin \Theta_{min}}\right)$, where $\Theta_{min}$ is the minimum interior angle of any face of $P$~\cite[Theorem 3.1]{lanthier_maheshwari_sack_01_weighted_geodesics}.

Aleksandrov et al.~\cite{aleksandrov_lanthier_maheshwari_sack_98_epsilonapproximation_geodesic} presented an algorithm to compute an approximation of a weighted geodesic path on arbitrary polyhedra with approximation ratio $(1+\epsilon)$. The algorithm is similar to~\cite{lanthier_maheshwari_sack_97_weighted_geodesics} in that Steiner points are added along each edge of $P$. On each edge, $m = O(\log\frac{L}{r})$ Steiner points are placed, where $L$ is the length of the longest edge of $P$ and $r$ is  $min(\frac{\epsilon}{2+3W/w}, \frac{1}{6})$ times the minimum distance of a vertex of $P$ to the boundary of the union of its incident faces. As before, $\frac{W}{w}$ is the ratio between the maximum and the minimum weight. A graph $G$ is computed on the Steiner points and $G$ is partitioned into $\frac{mn}{r}$ sub-regions. In each sub-region, all shortest paths between pairs of vertices are computed. Furthermore, all shortest paths between pairs of vertices on the boundaries of the sub-regions are computed. This computation takes $O(nmr \log r + \frac{(nm)^2}{r}\log\frac{nm}{\sqrt{r}}+ \frac{(nm)^2}{\sqrt{r}})$ time. The graph $G$ has complexity $O(nm^2)$, which dominates the space requirement of the algorithm. Note that since the graph is subdivided into small sub-graphs, the preprocessing is suitable for parallelizing the algorithm. After preprocessing, a $(1+\epsilon)$ approximation of the shortest path between two arbitrary query points on $P$ can be reported. To our knowledge, this algorithm has not been implemented.

Aleksandrov et al.~\cite{alksandrov_maheshwari_sack_00_approx_geodesics} extend this algorithm and place Steiner points on edges of $P$ and in the interior of faces of $P$. The approximation ratio remains $(1+\epsilon)$. That way, a graph $G$ is constructed. An extension of Dijkstra's algorithm can be run on $G$ to obtain a $(1+\epsilon)$ approximation of shortest paths on $P$. The algorithm takes $O(\frac{n}{\epsilon}\log\frac{1}{\epsilon}(\frac{1}{\sqrt{\epsilon}}+\log n))$, for $0<\epsilon<1$ and $O(n\log n)$ for $\epsilon \geq 1$ time to compute the shortest path between two arbitrary vertices on $P$. 

Sun and Reif~\cite{sun_reif_06_approximate_weighted} improve the algorithm by Aleksandrov et al.~\cite{alksandrov_maheshwari_sack_00_approx_geodesics} to run in $O(\frac{n}{\epsilon}\left(\log \frac{1}{\epsilon} + \log n \right) \log \frac{1}{\epsilon})$ time. This improvement is achieved by solving the SSSP problem on the graph enhanced by Steiner points using a new algorithm called \textit{Bushwhack algorithm}. The Bushwhack algorithm is similar to Dijkstra's algorithm. However, the Bushwhack algorithm maintains for each Steiner point a small set of incident edges that are likely to be used in order to improve the current shortest path. This list of edges results in an algorithm that is faster than Dijkstra's algorithm. Sun and Reif implemented and tested their algorithm. They found that when $O(\frac{1}{\epsilon}\log \frac{1}{\epsilon})$ Steiner points are inserted per edge, the hidden constant in the $O$-notation is large. 

Aleksandrov et al.~\cite{aleksandrov_maheshwari_sack_2003_approximate_geodesic, aleksandrov_maheshwari_sack_2005_approximate_geodesic} improve the running time of the algorithm to $O(\frac{n}{\sqrt{\epsilon}}\log\frac{n}{\epsilon}\log\frac{1}{\epsilon})$ by discretizing $P$ differently. In this algorithm, Steiner points are placed along the three bisectors of triangles of $P$. The practical use of the algorithms is limited, since a large number of Steiner points is inserted. Due to memory restrictions on current computers, this yields problems for real-life data sets.

If the weights used for the weighted distances are restricted to be in the range $\left[1, \rho\right]\cup\left\{\infty\right\},\rho \geq 1$, Cheng et al.~\cite{cheng_na_vigneron_wang_08_approximate} present an algorithm to compute an approximation of ratio $(1+\epsilon)$ of the shortest path from $s$ to $t$ that runs in $O(\frac{\rho^2 \log \rho}{\epsilon^2} n^3 \log \frac{\rho n}{\epsilon})$ time. The advantage of this algorithm is that the running time does not depend on the geometry of $P$.

Lanthier et al.~\cite{lanthier_nussbaum_sack_03_parallel_geodesics} implemented the first parallel algorithm to compute approximations of ratio $(1+\epsilon)$ for weighted shortest paths. As in previous approaches, the approach proceeds by constructing a graph and by computing the shortest path between vertices of a graph. The computation of the shortest paths is based on Dijkstra's algorithm and can be broken down into three components: preprocessing to find a graph $G$, executing Dijkstra's algorithm on $G$, and backtracking the path. The algorithm uses a spatial indexing structure called \textit{multidimensional fixed partition} that achieves load balancing and reduces the idle time of processors. The algorithm can solve the SSSP and the APSP problems. The algorithm was tested on a network of workstations, on a beowulf cluster, and on a symmetric multiprocessing architecture. The tests were performed for six geographic data sets with up to one million triangles and achieved acceptable running times.

Aleksandrov et al.~\cite{aleksandrov_djidjev_guo_maheshwari_sack_06_approx_geodesics_weighted} preprocess $P$ in $O\left(\frac{n}{\sqrt{\epsilon}} \log \frac{1}{\epsilon} \log\frac{n}{\epsilon}\right)$ time and $O\left(\frac{n}{\sqrt{\epsilon}} \log \frac{1}{\epsilon}\right)$ space, such that an approximation of ratio $(1+\epsilon)$ between any point $q$ on $P$ and a given source point $s$ on $P$ can be computed in $O\left(\frac{1}{\epsilon}\right)$ time. Alternatively, $P$ can be preprocessed in $O\left(\frac{(g+1)n^2}{\epsilon^{3/2}q}\log\frac{n}{\epsilon}\log^4\frac{1}{\epsilon}\right)$ time and $O\left(\frac{(g+1)n^2}{\epsilon^{3/2}q}\log^4\frac{1}{\epsilon}\right)$ space, such that an approximation of ratio $(1+\epsilon)$ between any pair of points on $P$ can be computed in $O(q)$ time, where $g$ is the genus of $P$ and where $q$ is an input parameter. The algorithm is complex and has not been implemented to our knowledge.

The following problems related to computing weighted shortest paths on the surface of a possibly non-convex polyhedron remain unsolved:
\begin{itemize}
\item No exact algorithm for computing weighted shortest paths exists to our knowledge.
\item How can $m$ Steiner points be placed on each face such that the best approximation accuracy is obtained~\cite{lanthier_99_phdThesis}? Is this problem NP-hard?
\item What is the minimum number of Steiner points needed on each face to obtain a $(1+\epsilon)$-approximation scheme? Is this problem NP-hard?
\end{itemize}

\section{Sample-Based Algorithms}
\label{smooth}
This section reviews algorithms for computing shortest paths on discretized smooth surfaces. We focus on the case where the discretization at hand is given as polyhedron. Unlike the above-mentioned algorithms, the algorithms reviewed in this section generalize algorithms from differential geometry to compute geodesic paths on smooth surfaces to operate on discretized surfaces. The research area concerned with these problems is \textit{discrete differential geometry}. For a more extensive survey, refer to Kirsanov~\cite{kirsanov_04_thesis}.

Kimmel and Kiryati~\cite{kimmel_kiryati_96_geodesics} assume that a discretized surface is given in a voxel representation. That is, space is divided into a cubical grid and each grid point is labeled as located inside the surface, on the surface, or outside of the surface. The approach proposed by Kimmel and Kiryati has two stages. In the first stage, a 3D length estimator is used in a graph search on the graph defined by the surface voxels to find a global approximation of the shortest path. This approximation is then refined using local information. The refinement is done using a discrete version of geodesic curvature shortening flow. This way, an approximation of a shortest path between two grid points can be found. The approximation ratio is not shown to be bounded. However, since the underlying surface is assumed to be smooth, the approximation is the best that can be obtained with the available voxel grid size. The algorithm has been implemented and tested thoroughly and appears to perform well in practice.

Kimmel and Sethian~\cite{kimmel_sethian_98_computing_geodesic} present an approach called fast marching method on triangular domains (FMM) that solves the SSSP problem by solving the Eikonal equation on a triangular grid with $n$ vertices. The result is based on Sethian's method to solve the Eikonal equation on a quadrilateral grid~\cite{sethian_96_fast_marching, Sethian99levelset}. The algorithm's running time is $O(n \log n)$. The algorithm proceeds by iteratively unfolding all of the triangles of the triangular mesh. When unfolding triangles, Steiner points are placed along the edges of the triangles. This results in shortest paths that cut through faces of the triangulation and yields consistent results. However, the shortest paths found using the FMM method only approximate the true geodesic distances on the triangular mesh. The reason is that the true geodesic distance may require Steiner points in the interior of a triangular face. The accuracy of the approach depends on the quality of the underlying triangulation; namely on the longest edge and the widest angle in the triangular mesh. The algorithm requires $O(n)$ space. Since the algorithm is easy to implement and performs well in practice, several implementations of the algorithm exist. Yatziv et al.~\cite{yatziv_bertesaghi_sapiro_06_fast_marching_linear} improve the running time of FMM by using an \textit{untidy priority queue}. Their experimental results show that the accuracy of the computed shortest paths only suffers slightly from this newly introduced inaccuracy. Kirsanov~\cite{kirsanov_04_thesis} introduces a novel update rule for FMM during the march. This update rule yields a higher accuracy of the resulting shortest paths. Bertelli et al.~\cite{bertelli_sumengen_manjunath_06_allPairs} consider solving the APSP problem using FMM. Their goal is to take advantage of the redundant computation in different passes of the SSSP algorithm to obtain a more efficient approach than simply running FMM $n$ times with each vertex as source point. Although the algorithm is shown to achieve higher efficiency in experiments, the worst case running time of the algorithm remains $O(n^2\log n)$.

Martinez et al.~\cite{martinez_velho_carvalho_04_geod} present a way to iteratively improve an existing estimate of a geodesic path between two vertices of a triangulated surface. Starting from a path computed via FMM, the path can be refined to yield a better approximation. Similar to Kimmel and Kiryati~\cite{kimmel_kiryati_96_geodesics}, a discrete geodesic curvature flow is used to iteratively improve the approximation. Martinez et al. show that the iterative scheme converges to the true geodesic path.

Xin and Wang~\cite{xin_wang_07_FMM} present another iterative method to improve the path found by the fast marching method. The algorithm first improves the initial fast marching method by classifying the edges of $P$ into different types and by treating different edge types differently during the wave front propagation. Second, the algorithm iteratively improves the resulting shortest path until the exact locally shortest path is found.

Memoli and Sapiro~\cite{memoli_sapiro_02_geodesics_pointcloud} approximate the geodesic distances of an underlying smooth manifold using a cloud of sample points without aiming to reconstruct a polyhedron representing the surface. The algorithm is based on a previous algorithm that operates on implicit surfaces. The algorithm proceeds by placing a ball around each sample point and by computing the union $U$ of those balls. The Euclidean distance in $U$ is used to approximate the geodesic distance on the underlying smooth manifold. The approximation is proven to be bounded by a constant if the sampling rate is sufficiently small. The sampling rate needed by the algorithm depends on the highest principal curvature of the underlying smooth surface. If the sampling is subject to noise, the bound on the approximation error gets worse. However, the decline in accuracy can be bounded by a function depending on the sampling noise.

\begin{table*}[h]
\begin{tabular}{|l|l|l|l|l|l|l|}\hline
& Polyhedral     & Cost  & Approximation & Time      & Ref.\\ 
& Surface        & Metric& Ratio         & Complexity&          \\ \hline\hline     

Graph-based & Convex & Euclidean & 1 & $O(n^3\log n)$      & \cite{sharir_schorr_86_geodesics_convex_polyhedron} \\ \hline
Graph-based & Convex & Euclidean & 1 & $O(n^2\log n)$ 		 & \cite{mount_85_geodesics_convex} \\ \hline
Graph-based & Convex & Euclidean & 2 & $O(n)$ 		 & \cite{hershberger_suri_95_geodesic_convex_approx} \\ \hline
Graph-based & Convex & Euclidean & $1+\epsilon$ & $O(n \min(\frac{1}{\epsilon^{1.5}}, \log n)+ \frac{1}{\epsilon^{4.5}} \log \frac{1}{\epsilon})$ 		 & \cite{har-peled_sharir_varadarajan_96_approximate_geodesics_convex} \\ \hline
Graph-based & Convex & Euclidean & $1+\epsilon$ & $O(n \log \frac{1}{\epsilon} + \frac{1}{\epsilon^3})$ 		 & \cite{agarwal_har-peled_sharir_varadarajan_97_approximating_geodesics} \\ \hline
Graph-based & Convex & Euclidean & 1 & $O(n\log n)$ 		 & \cite{schreiber_sharir_2006_optimal_geodesic_convex} \\ \hline \hline

Graph-based & Non-convex & Euclidean & 1 & $O(n^5)$ & \cite{orourke_suri_booth_85_geodesics} \\ \hline
Graph-based & Non-convex & Euclidean & 1 & $O(n^2\log n)$ & \cite{mitchell_mount_papadimitriou_87_geodesic} \\ \hline
Graph-based & Non-convex & Euclidean & 1 & $O(n^2)$ & \cite{chen_han_92_shortest_path} \\ \hline
Graph-based & Non-convex & Euclidean & $1+\epsilon$ & $O\left(n^2\log n+\frac{n}{\epsilon} \log \left(\frac{1}{\epsilon} \right) \log \left(\frac{n}{\epsilon} \right)\right)$ & \cite{har-peled_98_general} \\ \hline
Graph-based & Non-convex & Euclidean & $7(1+\epsilon)$ & $O(n^{\frac{5}{3}} \log^{\frac{5}{3}}n)$ & \cite{agarwal_varadarajan_97_approximate_geodesics} \\ \hline
Graph-based & Non-convex & Euclidean & $15(1+\epsilon)$ & $O(n^{\frac{8}{5}} \log^{\frac{8}{5}}n)$ & \cite{agarwal_varadarajan_97_approximate_geodesics} \\ \hline
Graph-based & Non-convex  & Euclidean   & 1         & $O(n\log^2n)$       & \cite{kapoor_99_geodesics} \\ \hline \hline

Graph-based & Non-convex  & Weighted & $1+\epsilon$ & $O(n^8 \log(nN\frac{W}{w\epsilon}))$ & \cite{mitchell_papadimitriou_87_weighted_geodesics} \\ \hline 
Graph-based & Non-convex & Weighted & Additive & $O(n^5)$ & \cite{lanthier_maheshwari_sack_01_weighted_geodesics}\\ \hline 
Graph-based & Non-convex & Weighted & $1+\epsilon$& $O(nmr \log r + \frac{(nm)^2}{r}\log\frac{nm}{\sqrt{r}}+ \frac{(nm)^2}{\sqrt{r}})$ & \cite{aleksandrov_lanthier_maheshwari_sack_98_epsilonapproximation_geodesic} \\ \hline
Graph-based & Non-convex & Weighted & $1+\epsilon$& $O(\frac{n}{\epsilon}\log\frac{1}{\epsilon}(\frac{1}{\sqrt{\epsilon}}+\log n))$ & \cite{alksandrov_maheshwari_sack_00_approx_geodesics} \\ \hline
Graph-based & Non-convex & Weighted & $1+\epsilon$& $O(\frac{n}{\sqrt{\epsilon}}\log\frac{n}{\epsilon}\log\frac{1}{\epsilon})$ & \cite{aleksandrov_maheshwari_sack_2005_approximate_geodesic} \\ \hline
Graph-based & Non-convex & Weighted & $1+\epsilon$& $O(\frac{n}{\epsilon}\left(\log \frac{1}{\epsilon} + \log n \right) \log \frac{1}{\epsilon})$ & \cite{sun_reif_06_approximate_weighted} \\ \hline
Graph-based & Non-convex & Weighted & $1+\epsilon$& $O(\frac{\rho^2 \log \rho}{\epsilon^2} n^3 \log \frac{\rho n}{\epsilon})$ & \cite{cheng_na_vigneron_wang_08_approximate} \\ \hline
Graph-based & Non-convex & Weighted & $1+\epsilon$& $O\left(\frac{n}{\sqrt{\epsilon}} \log \frac{1}{\epsilon} \log\frac{n}{\epsilon}\right)$ & \cite{aleksandrov_djidjev_guo_maheshwari_sack_06_approx_geodesics_weighted} \\ \hline \hline

Sample-based & Non-convex & Euclidean & Unbounded & $O(n\log n)$ &\cite{kimmel_sethian_98_computing_geodesic} \\ \hline
Sample-based & Non-convex & Euclidean & Unbounded & $O(n)$ &\cite{yatziv_bertesaghi_sapiro_06_fast_marching_linear} \\ \hline
\end{tabular} 
\caption{Results on Shortest Paths on a Polyhedral Surface $P$ with $n$ vertices. The constant $\epsilon > 0$ is the desired accuracy of the shortest path. In the weighted case, $N$ is the largest integer coordinate of any vertex in the subdivision and $\frac{W}{w}$ is the ratio between the maximum and the minimum weight. The symbol $m$ denotes the number of Steiner points placed along one edge. The symbol $r$ denotes $min(\frac{\epsilon}{2+3W/w}, \frac{1}{6})$ times the minimum distance of a vertex of $P$ to the boundary of the union of its incident faces. The constant $\rho > 1$ is the largest weight assigned to a face of $P$. }
 \label{table:results}
\end{table*}
\clearpage

The following problems related to sample-based geodesic computations remain unsolved:
\begin{itemize}
\item Graph-based algorithms find globally optimal paths that may not be locally optimal if the graph is based on samples obtained from a smooth surface. Algorithms from differential geometry can be discretized to find locally shortest paths. However, these algorithms can often get trapped in local insignificant minima. Can graph-based algorithms be combined with algorithms from discrete differential geometry to yield efficient globally convergent algorithms to compute a bounded approximation of the geodesic distance on a sample set obtained from a smooth surface~\cite{kirsanov_04_thesis}?
\item Can FMM be generalized to solve the APSP problem in $o(n^2\log n)$ time (recall that a solution in $O(n^2\log n)$ was suggested~\cite{bertelli_sumengen_manjunath_06_allPairs})?
\end{itemize}

\section{Summary}

To summarize this survey, Table \ref{table:results} gives the reviewed results on
shortest path problems on polyhedral surfaces.

\end{document}